\documentclass{elsart}

\begin{document}
\begin{frontmatter}

\title{Shape-phase transitions in mixed parity systems and the onset
of octupole deformation}
\author{Serdar Kuyucak\thanksref{email}}
\thanks[email]{E-mail: sek105@rsphysse.anu.edu.au}
\address{Department of Theoretical Physics,
Research School of Physical Sciences,\\
Australian National University, Canberra, ACT 0200, Australia}

\begin{abstract}
Shape-phase transitions in mixed parity boson systems are studied 
using mean field theory.  It is shown that without parity projection, 
shape transitions in mixed parity systems are very similar to those in 
positive parity systems, both requiring a finite interaction strength 
for the onset of deformation.  The shape-phase diagram in mixed parity 
systems, however, changes significantly after parity projection, with 
the critical point for the onset of octupole deformation moving to 
zero strength.  Shape transitions in the Lipkin model are shown to 
exhibit the same features, indicating that this is a direct 
consequence of parity projection in finite, mixed parity systems 
independent of the nature of interacting particles.
\\[3mm]

\end{abstract}

\end{frontmatter}

The low-lying negative parity states in even-even collective nuclei are 
described in terms of the octupole degree of freedom (see 
\cite{ahm93,but96} for recent reviews).  Experimental study of the 
octupole states is hindered by the much stronger quadrupole 
collectivity which limits their accessibility.  Thanks to the advances 
in detector technology, this situation has changed dramatically in 
recent years.  Exploiting the capabilities of the $\gamma$-ray arrays 
Eurogam and Gammasphere, the range of negative parity spectra in 
actinide nuclei Rn, Ra and Th has been greatly extended 
\cite{sim95,coc97}.  These new data, in turn, provide fresh challenges 
for collective models, the most interesting one being the description 
of the conjectured transition from the octupole-vibrational to the 
octupole-deformed shapes in actinides.  In this letter, we present 
mean field studies of this shape-phase transition in two analytically 
solvable models, the interacting boson model (IBM) \cite{iac87} and 
the Lipkin model \cite{lip65}, that will shed further light into this 
question.

We first briefly review the spherical $\to$ quadrupole-deformed shape 
transition in the $sd$-IBM \cite{die80,gin80,iac98}, which provides a 
useful reference point for the octupole case.  A convenient form of 
the Hamiltonian is
\begin{equation}
H_{sd}=  \varepsilon_d \hat n_d - \kappa_2  Q\cdot Q, 
\label{sdh}
\end{equation}
where the one-body term tries to keep the system spherical while the 
quadrupole interaction drives it towards deformation.  Here $\hat n_d= 
\sum_\mu d^\dagger_\mu d_\mu$ is the $d$-boson number operator and 
$Q = [s^\dagger \tilde d + d^\dagger s] +\chi_d [d^\dagger \tilde 
d]^{(2)}$ is the quadrupole operator.  For simplicity, we use the 
axially symmetric intrinsic state in the mean field analysis (the full 
state leads to the same result)
\begin{equation}
|N, \beta_2 \rangle = \left[N! (1+\beta_2^2)^N\right]^{-1/2} 
\left(s^\dagger + \beta_2 d_0^\dagger\right)^N |0 \rangle.
\label{con}
\end{equation}
Evaluating the expectation value of the Hamiltonian (\ref{sdh})
in the intrinsic state (\ref{con}) yields the energy surface
\begin{equation}
E(\beta_2)= N^2 \kappa_2 \left[ {\eta_2 \beta_2^2 \over 1+\beta_2^2}
-\left({ 2\beta_2 + \bar\chi_d \beta_2^2 \over 1+\beta_2^2}\right)^2 \right],
\label{es2} 
\end{equation}
where $\bar\chi_d=-\sqrt{2/7} \chi_d$, and
$\eta_2=\epsilon_d/N\kappa_2$ provides a convenient parameter to study
the shape transition.  Here and in the following, the $1/N$
corrections are neglected in the spirit of the large $N$ assumption
inherent in the mean field theory.  Variation of $E(\beta_2)$ with
respect to $\beta_2$ gives
\begin{equation}
\beta_2 \left[ \eta _2 (1+\beta_2^2)-2(2+\bar\chi_d 
\beta_2)(1+\bar\chi_d \beta_2 -\beta_2^2) \right]=0.
\label{var2}
\end{equation}
Analysis of Eqs.~(\ref{es2}, \ref{var2}) shows that the critical point 
for spherical $\to$ deformed shape transition is given by $\eta_2^{\rm 
cr} = 4+\bar \chi_d^2$.  For $\eta _2 > \eta_2^{\rm cr}$, the system 
has a spherical shape with $\beta_2=0$, and for $\eta _2 < \eta_2^{\rm 
cr}$, it becomes deformed with $\beta_2$ determined from 
Eq.~(\ref{var2}).  At the critical point, the spherical and deformed 
solutions coexist with energies $E(\beta_2=0)=E(\beta_2= \bar 
\chi_d/2)=0$.  We note that angular momentum projection leads to the 
same energy expression as (\ref{es2}) to leading order in $N$ 
\cite{kuy88}, hence restoration of the rotational invariance leads to 
a qualitatively similar shape-phase diagram.

The proper framework for discussion of shape-phase transitions in 
mixed parity systems is the $spdf$-IBM which has been shown to give an 
adequate description of the positive and negative parity spectra in collective 
nuclei \cite{eng85,ots86,ots88,kus88}.  However, the mean field 
analysis of the $spdf$-IBM is quite complicated which tends to obscure 
otherwise a very simple conclusion.  To illustrate the main ideas 
involved in a simple case, we first discuss the spherical $\to$ 
octupole-deformed shape transition in the $sf$-IBM using the 
Hamiltonian
\begin{equation}
H_{sf}=  \varepsilon_f \hat n_f - \kappa_3  O\cdot O,
\label{sfh}
\end{equation}
where $\hat n_f$ is the $f$-boson number and
$O = [s^\dagger \tilde f + f^\dagger s]^{(3)}$ 
is the octupole operator.  This Hamiltonian is similar to (\ref{sdh}) 
when $\chi_d=0$, and like the $\gamma$-instability in that case, it is 
unstable in the direction of deformation.  That is, each of the 
octupole mean fields $\beta_{3\mu}$ is populated equally, and the 
energy surface is a function of $\beta_3^2=\sum_\mu \beta_{3\mu}^2$.  
With this provision, the intrinsic state and the corresponding energy 
surface (without PP) are given by
\begin{eqnarray}
|N, \beta_3 \rangle &=& \left[N! (1+\beta_3^2)^N\right]^{-1/2} 
\left(s^\dagger + \vec \beta_3 \cdot \vec f^{\, \dagger}\right)^N |0 
\rangle, \nonumber \\ 
E(\beta_3) &=& N^2 \kappa_3 \left[ {\eta_3 \beta_3^2 \over 1+\beta_3^2}
-\left({ 2\beta_3 \over 1+\beta_3^2}\right)^2 \right],
\label{es3}
\end{eqnarray}
where $\eta_3=\epsilon_f/N\kappa_3$.
Variation of $E(\beta_3)$ w.r.t. $\beta_3$ gives
\begin{equation}
\beta_3 \left[\eta_3 (1+\beta_3^2)-4(1-\beta_3^2)\right]=0,
\label{var}
\end{equation}
which has the solutions 
\begin{eqnarray}
1) &&\ \beta_3 = 0, \quad E=0, \nonumber \\  
2) &&\ \beta_3^2 = {4-\eta_3 \over 4+\eta_3}, \quad 
E = - {N^2 \kappa_3 \over 16} (\eta_3-4)^2, \quad \eta _3 \le 4.
\label{sol}
\end{eqnarray}
It is clear from these solutions that for $\eta > 4$, the shape is 
spherical, and the critical point for the onset of octupole 
deformation occurs at $\eta_3^{\rm cr} = 4$.  Thus, just like in the 
quadrupole case, a finite interaction strength, given by $\kappa_3 = 
\epsilon_f /4N$, is required for the shape transition to take place.  
Note that one can obtain this result without explicit solutions by 
simply expanding the energy surface around $\beta_3$, and looking at 
the coefficient of $\beta_3^2$.  From Eq.~(\ref{es3}), $E(\beta_3) 
\simeq N^2 \kappa_3 [(\eta_3 -4)\beta_3^2 + \dots]$, and it is obvious 
from the sign of $\eta_3 -4$ that the $\beta_3 = 0$ solution is stable 
for $\eta_3 > 4$ but becomes unstable for $\eta_3 < 4$, leading to a 
deformed solution.  This method of determining the critical point is 
useful in cases where the extremum conditions cannot be solved 
explicitly.

The intrinsic state in Eq.~(\ref{es3}) has mixed parity, and it is 
important to see how PP influences the shape transition 
characteristics discussed above.  The  
projection operator onto good parity is given by
\begin{equation}
P_{\pi}=(1+\pi {\mathcal P})/2,
\label{pp}
\end{equation}
where $\pi=\pm 1$ is the parity quantum number and ${\mathcal P}$ is the 
parity operator.  Under ${\mathcal P}$, boson operators transform as
${\mathcal P} b^\dagger_{lm} {\mathcal P}=(-1)^l b^\dagger_{lm}$,
hence ${\mathcal P}$ acting on the intrinsic state (\ref{es3}) gives   
\begin{equation}
{\mathcal P} |N, \beta_3 \rangle = |N, -\beta_3 \rangle.
\label{pp1}
\end{equation}
From Eqs.~(\ref{pp}) and (\ref{pp1}), the projected states can be 
written as
\begin{equation}
P_{\pi} |N, \beta_3 \rangle = \left( |N, \beta_3 \rangle 
+\pi |N, -\beta_3 \rangle \right)/2.
\label{pps}
\end{equation}
Binomial expanding the even and odd spin parts in (\ref{pps}), it is 
easy to see that for $\pi=+1$, it has an even number of $f$ bosons, 
and for $\pi=-1$, it has an odd number of $f$ bosons.  
Normalization of the projected state (\ref{pps}) is given by
\begin{equation}
{\mathcal N}_\pi = \langle N, \beta_3 | P_{\pi} |N, \beta_3 \rangle 
=  (1+\pi r^N)/2,
\label{norm}
\end{equation}
where 
\begin{equation} 
r=(1-\beta_3^2)/ (1+\beta_3^2),
\label{r}
\end{equation}
gives a convenient measure of parity mixture in the condensate; for 
$r=1$ ($\beta_3=0$), it consists of only $s$ bosons, while for  
$r=0$ ($\beta_3=1$), it is an equal admixture of $s$ and $f$ bosons.

The energy surface of the Hamiltonian (\ref{sfh}) with PP is given by
\begin{eqnarray}
E_{\pi}(\beta_3) &=& {1 \over {\mathcal N}_\pi}  
\langle N, \beta_3 | H P_{\pi} |N, \beta_3 \rangle,  \nonumber \\
&=& {N^2 \kappa_3 \over 1+\pi r^N}
\left\{ {\eta_3 \beta_3^2 \over 1+\beta_3^2} \left(1-\pi r^{N-1}\right)
-\left({ 2\beta_3 \over 1+\beta_3^2}\right)^2 \right\}.
\label{es3p}
\end{eqnarray}
Variation of Eq.~(\ref{es3p}) leads to a complicated equation that 
cannot be solved analytically.  Nevertheless, by plotting the energy 
surface for various values of $\eta_3$, one can gain useful insights 
into the nature of the shape transition.  In Fig.~\ref{fig1} we show 
plots of Eqs.~(\ref{es3}) and (\ref{es3p}) for $\eta_3=4$, 40 and 400.  
The first value corresponds to the critical case without PP. The next 
two values represent a 10- and 100-fold reduction in the octupole 
strength compared to the first case for a fixed $\epsilon_f/N$.  In 
the top figure, the energy surface without PP (dashed line) exhibits 
the typical flat-bottomed curve expected at the critical point.  After 
PP the energy surface for $\eta_3=4$ is seen to become deformed (solid 
line).  What is surprising, however, is that despite the orders of 
magnitude reduction in the octupole strength in the middle and bottom 
figures, the $\beta_3=0$ solution remains unstable and the 
minimum of $E_+$ always occurs at $\beta_3^2>0$, that is, the energy 
surface after PP remains deformed.  As $\kappa_3$ is progressively 
reduced, the well depth and deformation decrease proportionally but 
the energy surface itself (solid line) stays more or less shape 
invariant.  To find the critical point analytically, we expand $E_+$ 
(\ref{es3p}) around $\beta_3=0$
\begin{equation}
E_+ \simeq N^2\kappa_3 \left[-2\beta_3^2 + (\eta_3 -2)\beta_3^4 + \dots\right].
\label{es3+}
\end{equation}
It is clear from Eq.~(\ref{es3+}) that for any $\kappa_3>0$, there is 
a deformed minimum, and hence the critical point is given by 
$\kappa_3=0$.  That is, the onset of the octupole deformation is 
immediate in the presence of any octupole strength, and the spherical 
phase found in the study of shape transition without PP disappears 
after PP. This situation is depicted in Fig.~\ref{fig2}, where 
$\beta_3$ corresponding to the absolute minimum of the energy surface 
with and without PP are plotted against $1/\eta_3$.  The sharp 
spherical $\to$ octupole-deformed transition exemplified by the dashed 
curve (which is also representative of the quadrupole case) is 
replaced by a smoothly varying curve after PP, and the critical point 
moves from $1/\eta_3=0.25$ to 0 (or from $\kappa_3=\epsilon_f/4N$ to 0).

To understand this somewhat surprising result, it is worthwhile to 
look at the effect of PP on individual terms by comparing 
Eqs.~(\ref{es3}) and (\ref{es3+}).  The leading order ($\beta_3^2$) 
contribution to the one-body energy is seen to vanish after PP, 
whereas the octupole term survives, loosing only half of its strength.  
A more direct way to see the effect of PP is to expand the condensate 
\begin{eqnarray}
\left(s^\dagger + \vec \beta_3 \cdot \vec f^{\, \dagger}\right)^N 
&=& (s^\dagger)^N + N (s^\dagger)^{N-1} \vec \beta_3 \cdot \vec f^{\, \dagger}
\nonumber \\
&&+ (1/2) N(N-1) (s^\dagger)^{N-2} (\vec \beta_3 \cdot 
\vec f^{\, \dagger})^2 + \dots ,
\end{eqnarray}
and look at the overlaps of various terms.  Denoting the states in 
the expansion by their $f$ boson number $n_f$, the leading 
contribution to the one-body term comes from the (1-1) matrix element 
(m.e.)  (i.e. $\langle 1 | \hat n_f | 1 \rangle$), which goes like 
$\beta_3^2$.  However, after PP, odd $n_f$ terms are projected out, so 
the leading term has to come from the (2-2) m.e.,  which goes like 
$\beta_3^4$.  In contrast, the leading (equal) contributions to the 
octupole term come from the (1-1) and (0-2 + 2-0) m.e., and PP blocks 
only the first one, leaving the second one intact.  Thus the net 
effect of PP is to lift the obstruction of the one-body operator 
against the formation of a deformed system.

The above conclusion on the effect of PP on shape transitions in mixed
parity systems can be generalized to arbitrary boson spins, and in
particular to the $spdf$-IBM. However, as stressed before this case is
rather complicated and is deferred to a more detailed study.  Due to
weaker dipole collectivity, the $p$ boson is weakly coupled to the
$sdf$ system, and should only have a perturbative effect on the
$sdf$-IBM results.  Thus, for our purposes, it is sufficient to study
the physically most relevant case, namely, the onset of octupole
deformation in a quadrupole deformed system within the framework of
the $sdf$-IBM. The Hamiltonian is given by a combination of
Eqs.~(\ref{sdh}) and (\ref{sfh})
\begin{equation}
H_{sdf}= \varepsilon_d \hat n_d + \varepsilon_f \hat n_f
-\kappa_2 Q\cdot Q - \kappa_3 O.O,
\label{hamsdf}
\end{equation}
where the quadrupole and octupole operators are generalized to
\begin{eqnarray}
Q &=& [s^\dagger \tilde d + d^\dagger s]^{(2)} 
+\chi_d [d^\dagger \tilde d]^{(2)} 
+\chi_f [f^\dagger \tilde f]^{(2)}, \nonumber \\
O &=& [s^\dagger \tilde f + f^\dagger s]^{(3)} 
+\chi_3 [d^\dagger \tilde f+f^\dagger \tilde d]^{(3)}. 
\end{eqnarray}
Numerical studies of this Hamiltonian with a general intrinsic state 
show that the ground state remains axially symmetric for realistic 
choices of parameter \cite{kuy99}.  
Using the intrinsic state
\begin{equation}
|N, \beta_2, \beta_3 \rangle =
{\left(s^\dagger + \beta_2 d_0^\dagger + \beta_3 f_0^\dagger\right)^N 
\over  \left[N! (1+\beta_2^2+\beta_3^2)^N\right]^{1/2} }|0 \rangle,
\label{consdf}
\end{equation}
the energy surface of the Hamiltonian (\ref{hamsdf}) after PP is given by 
\begin{eqnarray}
E_{\pi}(\beta_2,\beta_3) &=& {N^2 \over 1+\pi r^N} 
\Biggr\{ \sum_{l=2}^3 {\epsilon_l \beta_l^2 \over N \beta_+^2} 
\left(1+(-1)^l \pi r^{N-1}\right) \nonumber \\
&&-{\kappa_2 \over \beta_+^4}\left[ (2 \beta_2 + \bar\chi_d \beta_2^2 
+ \bar\chi_f \beta_3^2)^2  + \pi r^{N-2} (2 \beta_2 
+ \bar\chi_d \beta_2^2 - \bar\chi_f \beta_3^2)^2 \right] 
\nonumber \\
&& - {\kappa_3 \over \beta_+^4} \left( 2\beta_3  
+ 2\bar\chi_3 \beta_2 \beta_3 \right)^2 \Biggr\},
\label{espp}
\end{eqnarray}
where $\beta_+^2 = 1+\beta_2^2 +\beta_3^2$, \ $r=(1+\beta_2^2 
-\beta_3^2)/\beta_+^2$, $\bar\chi_f= \sqrt{4/21} \chi_f$ and 
$\bar\chi_3= -\sqrt{4/15}\chi_3$.  
The energy surface without PP follows from Eq.~(\ref{espp}) by setting 
$\pi=0$.  In the case of no PP, an analysis similar to 
Eqs.~(\ref{var}, \ref{sol}) can be carried out leading to an octupole 
vibrational phase ($\beta_3=0$ and $\beta_2$ given by 
Eq.~(\ref{var2})), and a deformed phase ($\beta_3>0$) which is too 
complicated to present here.  Nevertheless the critical point can be 
obtained from the expansion of the energy surface around $\beta_3=0$ 
as
\begin{equation}
\kappa_3 = { \epsilon_f (1+\beta_2^2)
+ 2N \kappa_2 \beta_2 (2+\bar\chi_d \beta_2) 
(\beta_2-\bar\chi_f)   \over 4N(1+\bar\chi_3 \beta_2)^2}.
\label{sdfcr}
\end{equation}
The quadrupole deformation is seen to have modified the critical point 
for the onset of octupole deformation but still a finite $\kappa_3$ is 
required.  A similar expansion of Eq.~(\ref{espp}) for $\pi=+1$ shows 
that the leading term is proportional to $-\kappa_3 \beta_3^2$  as 
in the spherical case (\ref{es3+}).  Thus after PP, the critical point 
again occurs at $\kappa_3=0$.

As a final example, we consider shape transitions in the Lipkin model 
\cite{lip65}, which has been shown to mimic, at the mean field level, 
the octupole deformation effects obtained from the more realistic 
Hartree-Fock-Bogolyubov calculations \cite{rob92}.  In the Lipkin 
model, $N=2j+1$ nucleons occupy two levels that have the same angular 
momentum $j$ but opposite parity $\pi$.  The Hamiltonian with one- and 
two-body terms is given by
\begin{equation}
H = \epsilon K_0 - {1\over 2} \kappa (K_+ K_+ + K_- K_- ),
\label{liph}
\end{equation}
where $K$'s are the SU(2) quasi-spin operators defined as
\begin{eqnarray}
K_0 &=& {1\over 2}\sum_{m} \left( a^\dagger_{m+} a_{m+} 
- a^\dagger_{m-} a_{m-} \right), \nonumber \\
K_+ &=& \sum_{m} a^\dagger_{m+}  a_{m-}, \quad K_- = (K_+)^\dagger,
\label{k}
\end{eqnarray}
with the quantum numbers $k=N/2$, $\mu=-N/2, 
-N/2+1, \dots, N/2$.

In the ``spherical'' (or harmonic) limit ($\kappa=0$), the ground 
state and its energy are given by
\begin{equation}
|\phi_0 \rangle = \prod_m a^\dagger_{m-} |0\rangle,  
\quad E_0 = -\epsilon N/2.
\end{equation}
The excited states are obtained by repeated application of $K_+$ on 
the ground state, which generates a finite, harmonic spectrum, $E_n = 
E_0 + n\epsilon, \ n=1, \dots,N$.  For $\kappa>0$, the two 
body-interaction mixes the $\pi=+1$ and $-1$ levels, and to find the 
ground state of the system, one needs to use an intrinsic state with 
mixed parity
\begin{equation}
|N, \beta \rangle = \left[(1+\beta^2)^N\right]^{-1/2} \prod_m 
\left( a^\dagger_{m-} +\beta a^\dagger_{m+} \right) |0\rangle,
\label{lips}
\end{equation}
where the mean field $\beta$ describes the ``deformation'' of the 
system.  With the intrinsic state (\ref{lips}), the energy surface of 
the Hamiltonian (\ref{liph}) without PP is given by
\begin{equation}
E(\beta) = -{\epsilon N\over 2} + N^2 \kappa \left[ 
{\eta \beta^2 \over 1+\beta^2}
-\left({ \beta \over 1+\beta^2}\right)^2 \right],
\label{lipe}
\end{equation}
where $\eta=\epsilon/N\kappa$.  Apart from the constant term, the 
energy surface (\ref{lipe}) is very similar to that of the $sf$-IBM 
(\ref{es3}), and it is easy to see from its expansion that the 
critical point for the onset of deformation occurs at $\eta = 1$ or 
$\kappa = \epsilon/N$.

Parity projection in the Lipkin model is defined just like in the 
boson case Eqs.~(\ref{pp}-\ref{pps}) and yields the following energy 
surface
\begin{equation}
E_{\pi}(\beta) = \left[-{\epsilon Nr\over 2}
-N^2 \kappa \left({ \beta \over 1+\beta^2}\right)^2 \right]
{ 1+\pi r^{N-2} \over 1+\pi r^N}, 
\end{equation}
where $r$ is defined as in Eq.~(\ref{r}).  Expanding $E_{+}(\beta)$ 
around $\beta=0$ gives
$E_+ \simeq -\epsilon N/2 - N^2 \kappa \beta^2 + \dots$.
Thus the critical point again moves to $\kappa=0$ after PP.

In conclusion, using analytically solvable models of mixed parity
systems we have demonstrated that a sharp transition to octupole
deformation, similar to the onset of quadrupole deformation, occurs
only if one ignores PP. After PP, the character of the shape
transition changes completely, and the vibrational phase disappears. 
While one can still use the terminology and formalism of octupole
vibrations for small $\beta_3$, this is only an approximation and not
a well defined phase as in quadrupole vibrations.  A similar effect of
the restoration of discrete symmetries in finite systems is well known
in the BCS theory, where the critical pairing strength vanishes after
the number projection.  The effect of PP on shape transitions, however,
seems to have been ignored in the literature so far.  Implications of
this result for the octupole states in actinides will be discussed in
a longer article.

Parts of this work were carried out while visiting the Japan Atomic Energy 
Research Institute and the Institute for Nuclear Theory at the 
University of Washington.  Hospitality and support of these Institutes 
are acknowledged with thanks.

% \vfill
% \eject

{\bf Figure Captions}
\\[1mm]
\begin{figure}[!h]
\caption{Energy surface with $\pi=+1$ (solid line) and without PP
(dashed line) for various values of $\eta_3$ ranging from 4 to 400. 
The former corresponds to the critical value without PP.
The energy surface with $\pi=-1$ is also shown in the top figure
(dotted line) but it is too high in energy to show in the other
cases.}
\label{fig1}
\end{figure}

\begin{figure}[!h]
\caption{Shape-phase diagram for spherical $\to$ octupole deformation 
with PP (solid line) and without PP (dashed line).  Deformations, 
obtained from the minima of the energy surfaces Eqs.~(\protect\ref{es3}) and
(\protect\ref{es3p}), are plotted against $1/\eta_3=N\kappa_3/\epsilon_f$.}
\label{fig2}
\end{figure}

\vfill

\begin{thebibliography}{9}
\bibitem{ahm93} I. Ahmad and P.A. Butler, Ann. Rev. Nucl. Part. Sci. 
{\bf 43} (1993) 71.  
\bibitem{but96} P.A. Butler and W. Nazarewicz, 
Rev. Mod. Phys. {\bf 68} (1996) 349.  
\bibitem{sim95} J.F. Smith et al., Phys. Rev. Lett. {\bf 75} (1995) 1050.  
\bibitem{coc97} J.F.C. Cocks et al., Phys. Rev. Lett. {\bf 78} (1997) 292;
Nucl. Phys. A {\bf 645} (1999) 61.  
\bibitem{iac87} F. Iachello and A. Arima,  The Interacting Boson Model
(Cambridge University Press, Cambridge, 1987). 
\bibitem{lip65} H.J. Lipkin, N. Meshkov and A.J. Glick, 
Nucl. Phys. A {\bf 62} (1965) 188.
\bibitem{die80} A.E.L. Dieperink and O. Scholten, 
Nucl. Phys. A {\bf 346} (1980) 125.
\bibitem{gin80} J.N. Ginocchio and M.W. Kirson, 
Nucl. Phys. A {\bf 350} (1980) 31. 
\bibitem{iac98} F. Iachello, N.V. Zamfir and R.F. Casten, 
Phys. Rev. Lett. {\bf 81} (1998) 1191.
\bibitem{kuy88} S. Kuyucak and I. Morrison, Ann. Phys. (NY) 
{\bf 181} (1988) 79.
\bibitem{eng85} J. Engel and F. Iachello, Phys. Rev. Lett. 
{\bf 54} (1985) 1126; Nucl. Phys. A {\bf 472} (1987) 61.  
\bibitem{ots86} T. Otsuka, Phys. Lett. B {\bf 182} (1986) 256.  
\bibitem{ots88} T. Otsuka and M. Sugita, Phys. Lett. B {\bf 209} (1988) 140.  
\bibitem{kus88} D. Kusnezov and F. Iachello, 
Phys. Lett. B {\bf 209} (1988) 420.
\bibitem{kuy99} S. Kuyucak and M. Honma, to be published.
\bibitem{rob92} L.M. Robledo, Phys. Rev. C {\bf 46} (1992) 238.
\end{thebibliography}
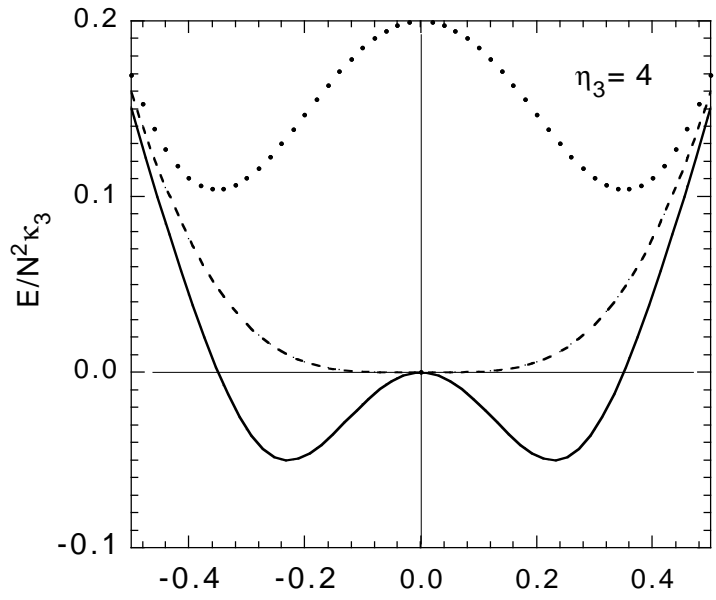
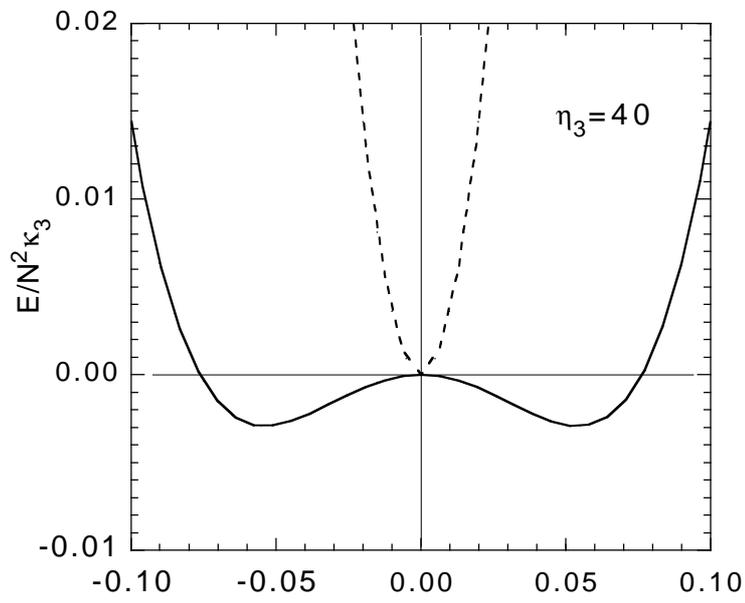
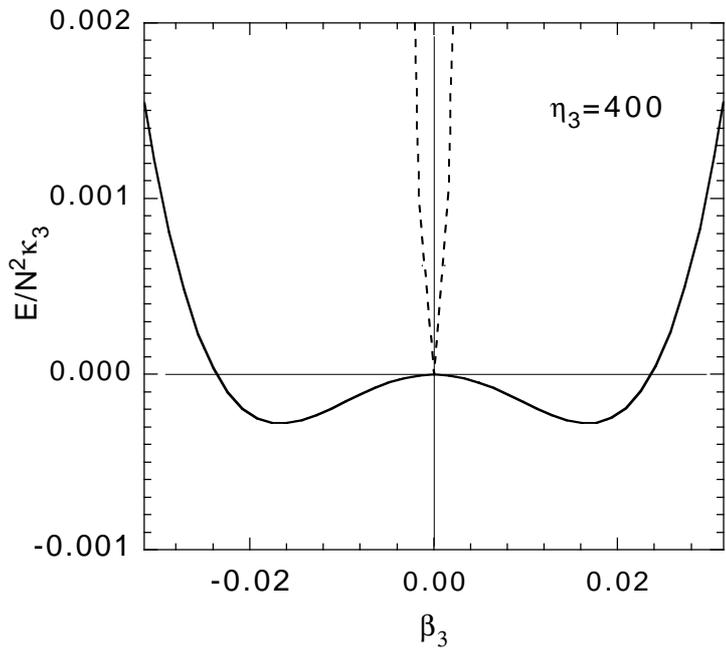
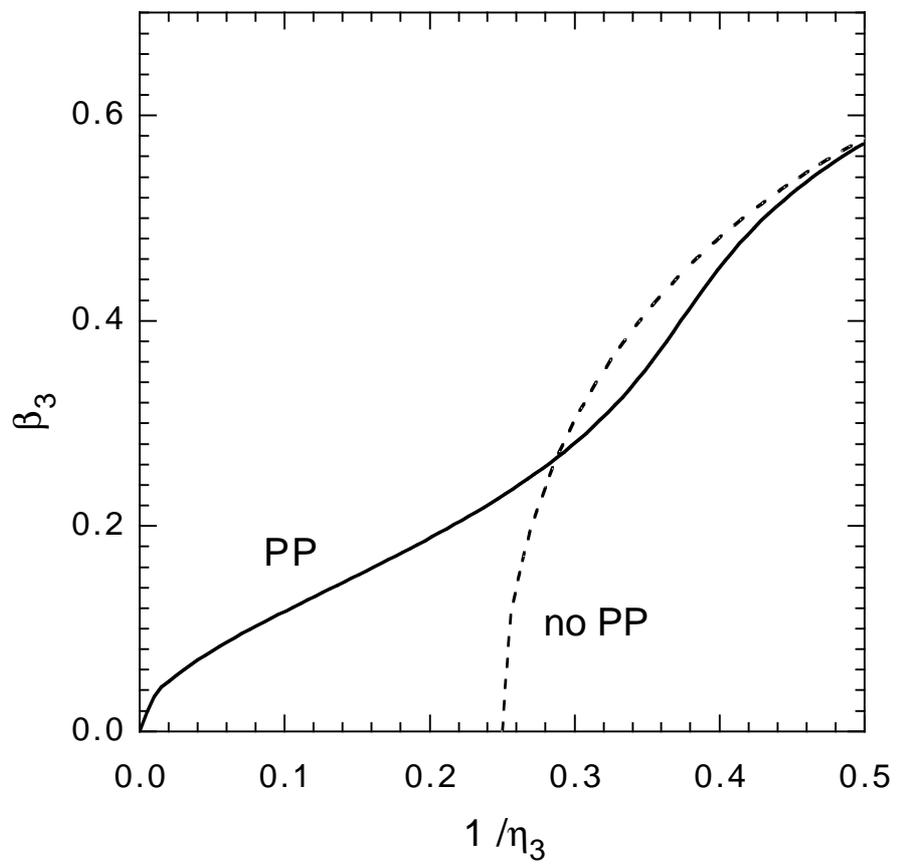
\end{document}